\documentclass[12pt]{article}

\usepackage{epsfig}
\usepackage{graphicx}
\usepackage{psfrag}
\usepackage{amssymb}
\usepackage{amsmath}

\newcommand{\rf}[1]{(\ref{#1})}
\newcommand{\beq}{\begin{equation}}

\newcommand{\eeq}{\end{equation}}
\newcommand{\bea}{\begin{eqnarray}}
\newcommand{\eea}{\end{eqnarray}}

\newcommand{\e}{\mbox{e}}
\renewcommand{\d}{\mbox{d}}

\newcommand{\lam}{\lambda}

\newcommand{\om}{\omega}

\newcommand{\del}{\delta}
\newcommand{\Del}{\Delta}

\newcommand{\kp}{\kappa}

\newcommand{\oh}{\frac{1}{2}}

\newcommand{\ra}{\rangle}
\newcommand{\la}{\langle}

\newcommand{\cD}{{\cal D}}

\newcommand{\cO}{{\cal O}}

\begin{document}

\begin{center}
\vspace{24pt}
{ \large \bf Renormalization in quantum theories of geometry}

\vspace{24pt}

{\sl J. Ambjorn}$\,^{a,b}$,
{\sl J. Gizbert-Studnicki}$\,^{c}$
{\sl A. G\"{o}rlich}$\,^{c}$,\\
{\sl J. Jurkiewicz}$\,^{c}$ and
{\sl R. Loll}$\,^{b}$

\vspace{10pt}

{\small

$^a$~The Niels Bohr Institute, Copenhagen University\\
Blegdamsvej 17, DK-2100 Copenhagen \O , Denmark.\\
email: ambjorn@nbi.dk

\vspace{10pt}

$^b$~Institute for Mathematics, Astrophysics and Particle Physics
(IMAPP)\\ Radboud University Nijmegen, Heyendaalseweg 135, \\
6525 AJ  Nijmegen, The Netherlands\\
email: r.loll@science.ru.nl, j.ambjorn@science.ru.nl

\vspace{10pt}

$^c$~Institute of Theoretical Physics, Jagiellonian University,\\
 \L ojasiewicza 11, Krak\'ow, PL 30-348, Poland.\\
email: jakub.gizbert-studnicki@uj.edu.pl, andrzej.goerlich@uj.edu.pl, jerzy.jurkiewicz@uj.edu.pl.

}

\end{center}

\vspace{24pt}

\begin{center}
{\bf Abstract}
\end{center}

\noindent
A hallmark of non-perturbative theories of quantum gravity is the absence of a fixed background geometry, and therefore the absence in a Planckian regime of any notion of length or scale that is defined a priori. This has potentially far-reaching consequences for the application of renormalization group methods \`a la Wilson, which rely on these notions in a crucial way. We review the status quo of attempts in the Causal Dynamical Triangulations (CDT) approach to quantum gravity 
to find an ultraviolet fixed point associated with the second-order phase transitions observed in the lattice theory.
Measurements of the only invariant correlator currently accessible, that of the total spatial three-volume, has not produced any evidence of such a fixed point. A possible explanation for this result is our incomplete and perhaps na\"ive understanding of what constitutes an appropriate notion of (quantum) length near the Planck scale.

\newpage

\section{Introduction}
\label{intro}

The Wilsonian concept of renormalization has been of immense importance for our understanding 
of quantum field theory and its relation to critical phenomena in statistical mechanics and condensed matter 
physics. In the context of lattice field theory it has been the guiding principle for approaching a continuum 
quantum field theory, starting out with a lattice regularization of the theory. Usually we view the 
ultraviolet (UV) regularization of the quantum field theory as a step on the way to defining the theory. For a given 
theory there will in general be many ways to introduce such a regularization, some more convenient than 
others, depending on the calculations one wants to perform. The lattice regularization is usually not the 
most convenient regularization if one wants to perform analytic calculations, but for some theories it allows
one to perform non-perturbative calculations, for instance in the form of Monte Carlo (MC) simulations of the 
field theories in question. It also allows one to address in a non-perturbative way the question of whether or not
a given quantum field theory exists, the simplest example being a $\phi^4$-theory in four dimensions.
This is a perturbatively renormalizable quantum field theory, so one can fix the physical  mass and 
the physical  coupling constant of the theory, and to any finite order in the coupling constant 
calculate the correlation functions. However, this does not imply that the theory really exists in the 
limit where the UV cut-off is taken to zero, since the perturbative expansion is only an asymptotic 
expansion. The lattice field formulation of the $\phi^4$-theory provides us with a tool to go beyond perturbation
theory, and (as will be discussed below) the result is that the $\phi^4$-quantum field theory does not exist
in four spacetime dimensions. In a similar vain, lattice field theory seems to confirm 
the existence of the quantum version of non-Abelian gauge theories.

The lattice field theories address the question of existence of certain quantum field theories using the 
Wilsonian picture: if the continuum quantum field theory exists as a limit of the lattice field theory when the cut-off 
is removed (the lattice spacing goes to zero), there exists a UV fixed point of the renormalization 
group.  One can approach such a fixed point in the following way: choose  observables which define the 
physical coupling constants of  the theory
and measure them for a certain choice of the {\it bare} coupling constants used to define the lattice theory.
Then change the lattice spacing by a factor 1/2 and find the new bare coupling constants which leave the 
observables unchanged\footnote{Using a description like this we assume we are so close to a continuum limit 
that  we can use a continuum language for the observables, rather than referring explicitly to the lattice. In addition,
note that 
a procedure like this will not leave all observables unchanged, but only the physical coupling constants. 
One could follow another renormalization procedure, where the action contained ``all possible coupling 
constants''. In this space one could follow a path which leaves all observables invariant.}. Continue halving the lattice spacing and in this way create  a flow of the 
bare coupling constants. The bare coupling constants will then flow to a UV fixed point (if it exists).

The next question is which observables to choose. In the case of a $\phi^4$-theory this is simple (and we will 
make a choice below). In the case of non-Abelian gauge theories it is already somewhat more difficult,
since observables should be gauge-invariant, while the theory is usually not formulated in terms of gauge-invariant 
variables. In MC simulations of the quantum field theory 
it is important to choose such gauge-invariant observables, since in quantum field theories the 
quantum fluctuations are dominated by UV fluctuations. If one uses the path integral (as one does 
in MC simulations), it implies that a typical field configuration is almost nothing but UV fluctuations.
This is true also for scalar theories like a $\phi^4$-theory, but since the field variables  in gauge theories 
are not gauge-invariant, most of these fluctuations are even more unphysical ``noise''. However, this noise will cancel 
when calculating expectation values of gauge-invariant observables. If we next move to quantum theories
of geometry, in particular attempts to quantize General Relativity (GR), the choice 
of  ``gauge-invariant'' observables becomes even more tricky. Gauge invariance in this context is usually replaced by 
diffeomorphism invariance, and there are few  invariant local observables. However, it is even more 
important that the concept of ``distance'' now becomes field-dependent. For a given geometry the 
distance between two points depends on the geometry. Therefore, if  we integrate over 
geometries  in the path integral, it becomes unclear how to think about a quantum correlation 
between fields  as function of a distance.
In particular, since distance, or scale, is paramount in the Wilsonian theory of critical phenomena, a new 
challenge arises  in this program when we  quantize geometries. 
This is what we want to discuss in this article. 

In Sec.\ \ref{sec2} we review the standard Wilsonian picture for a $\phi^4$-theory in four flat 
spacetime dimensions, emphasizing how to find a UV fixed point in the {\it bare} coupling constant space  
of the theory. In Sec.\ \ref{sec3} we discuss how to use the Wilsonian picture for the theory of quantum geometry
denoted Causal Dynamical Triangulations (CDT), which has been suggested as a theory of quantum gravity.     
Sec.\ \ref{sec4} discusses some examples where ``quantum distances'' appear in correlation functions,
whether these distances are observables and to what extent  the ``fractal structure'' of quantum geometry 
can be observed. Finally, Sec.\ \ref{sec5} contains a discussion.

\section{Approaching a UV fixed point}\label{sec2}

Let us consider a $\phi^4$-field theory on a four-dimensional hypercubic  lattice with periodic boundary conditions.
We assume that  the lattice
has $L_1$, $L_2$, $L_3$ and $L_4$ lattice links in the four directions, and that $L_i \gg 1$. The total 
number of lattice points is $N = L_1\cdots L_4$. If  the lattice spacing is $a_0$,
the corresponding physical volume is $V = N a_0^4$. Let $n =(n^1,\ldots,n^4)$ denote 
the integer lattice coordinates of the vertices. The corresponding spacetime coordinates will be $x_n=a_0 n$.
A scalar field $\phi_0$ lives on the lattice vertices and we write $\phi_0(n)$ or $\phi_0(x_n)$.
The lattice field theory action is 
\beq\label{s2-3}
S[\phi_0,m_0,\lam_0;a_0] = \sum_{n} a_0^4 \Big( \oh\sum_{i=1} ^4\frac{(\phi_0(n\!+\!\hat{i}) \!-\!\phi_0(n))^2}{a_0^2} + 
\oh m_0^2\phi_0^2(n) +\frac{1}{4!}\lam_0 \phi_0^4(n)\Big),
\eeq
where $\hat{i}$ denotes a unit vector in direction $i$.
The action is characterized by two so-called ``bare'', dimensionless  coupling constants $m_0a_0$ and $\lam_0$.
A correlation function is  defined as 
\beq\label{s2-4}
\la \phi_0 (n_1) \cdots \phi_0 (n_k) \ra = 
 \frac{ \int \prod_n d\phi_0(n) \;\phi_0 (n_1) \cdots \phi_0 (n_k) \e^{-S[\phi_0,m_0,\lam_0;a_0] } }{ 
 \int \prod_n d\phi_0(n)  \e^{-S[\phi_0,m_0,\lam_0;a_0] }}.
 \eeq
 We obtain the same action if we  simultaneously change $a_0 \to a$, set $a_0\phi_0 =a \phi$, $m_0 a_0 =m a$
and leave $\lam_0$ unchanged, and we have trivially 
\beq\label{s2-5}
 a^k_0\la \phi_0 (n_1) \cdots \phi_0 (n_k) \ra_{a_0,m_0,\lam_0} = a^k\la \phi (n_1) \cdots \phi (n_k) \ra_{a,m,\lam_0} .
\eeq

In the theory we also have   
renormalized coupling constants $m_R$ and $\lam_R$, which are determined 
by some explicit prescription, allowing us to ``measure'' them. For instance, $m_R$ can be defined 
from the exponential fall-off of the two-point function, while $\lambda_R$ can be defined as the 
connected four-point function at zero momentum. We thus have $m_R a_0 = 1/\xi$,
where $\xi$ is the dimensionless correlation length of the two-point $\phi$-correlator, 
measured in units of the lattice spacing. Similarly, there is an explicit definition of $\lam_R$.
Let us state how to measure these quantities on the lattice since we will use the same 
techniques in the case of gravity. We choose one of the lattice axes as the ``time" direction and define
the spatial average
\beq\label{s2-20}
\Phi_0(n^t) := \sum_{\vec{n}}  \phi_0 (\vec{n},n^t),\qquad \vec{n} = (n^x,n^y,n^z),
\eeq
and we have 
\beq\label{s2-21}
\la \Phi_0(n^t_1) \Phi_0(n_2^t)\ra_c  =  { \it const.} \; \e^{-|n^t_1 -n^t_2|/ \xi} + \cdots,
\eeq
where the subscript $c$ in $\la \cdot\ra_c$ is the connected part, and  
the dots indicate terms falling off faster at large time differences. The exponential decay for large
$|n^t_1-n_2^t|$ determines the physical mass $m_R  = 1/(a_0 \xi)$.
Similarly, we can define the susceptibilities
\beq\label{s2-22}
\chi_k := \sum_{n^t_1,\ldots,n^t_{k-1}} \la \Phi_0 (n_1^t)\cdots \Phi_0(n_{k-1}^t)\Phi_0(0)\ra_c
\eeq 
and the second moment
\beq\label{s2-23}
\mu_2 := \;\sum_{n^t} (n^t)^2 \la \Phi_0(n^t) \Phi_0(0)\ra_c.
\eeq
One then obtains\footnote{For a detailed discussion see the book \cite{munster}.}   (in the case $\la \phi_0(n)\ra =0$
 where there is no symmetry breaking) 
\beq\label{s2-24}
\lam_R = - \frac{16 \chi_4}{\mu_2^2}.
\eeq
$a_0$ is a fictitious parameter in the above formulation in the sense 
that if we make the above-mentioned change from $(a_0,\phi_0,m_0,\lam_0)$ to $(a,\phi,m,\lam_0)$ we obtain
the same $\xi$ and the same $\lam_R$, while $m_R$ changes in a trivial way since $\xi$ is unchanged.

 Let us choose a  value for $\lam_R$. 
For given values $(m_0a_0,\lam_0)$ of the bare coupling constants  we obtain a value $\lam_R(m_0a_0,\lam_0)$.
Among these there will be sets  $(m_0(s)a_0,\lam_0(s))$, parametrized by some parameter $s$, 
such that $\lam_R(m_0(s)a_0,\lam_0(s))=\lam_R$. They form a curve in the $(m_0a_0,\lam_0)$-coupling 
constant space. Note that this curve is unchanged if we change  $a_0 \to a$ and $m_0 \to m = m_0 a_0/a$ and 
consider the $(m a, \lam_0)$-coupling constant plane.   
Moving along this curve, the correlation length $\xi(s)$ will  change, so we can 
exchange our arbitrarily chosen parameter $s$ with $\xi$. If we reach a point along the curve where 
$\xi = \infty$, we have reached a second-order phase transition point  in the $(m_0a_0,\lam_0)$-coupling 
constant plane. This point can now serve as a UV fixed point for the $\phi^4$-theory, since we are 
free to insist that $m_R$ is constant along the curve provided that we redefine $a$ such  that 
$m_R a(\xi) = 1/\xi$. This will define $a(\xi)$ as a function of $\xi$, and -- since we are 
free to define the lattice theory with $a(\xi)$ instead of $a_0$ -- if we at the same time make a trivial 
rescaling of $m_0$ to $m(\xi) =m_0a_0/a(\xi)$, we will in this redefined theory obtain the same $\xi$ and $\lam_R$.
Thus it can be viewed as a rescaling of the lattice to smaller  $a$ while keeping the continuum physics 
(i.e. $m_R$ and $\lam_R$) constant. In particular, the correlation length in real spacetime is kept fixed
since $ |x|_{\rm corr} \equiv \xi a(\xi) = 1/m_R$. 
In the limit $\xi \to \infty$ the lattice spacing goes to zero and 
we have our continuum quantum field theory with the cut-off removed. 

The approach to  this assumed UV fixed point is governed by the so-called $\beta$-function\footnote{A priori
the $\beta$-function is a function of $\lam_0$ and $m_0a_0$, but one can show that close to the fixed point
one can ignore the $m_0a_0$-dependence.}, 
which relates the change in $\lam_0$ to the change in $a(\xi) = 1/(m_R \xi)$ as we move along
the trajectory of constant $m_R,\lam_R$,
\beq\label{s2-1}
-a \frac{\d \lam_0}{\d a}\Big|_{m_R,\lam_R} = \xi \frac{\d \lam_0}{\d \xi}\Big|_{m_R,\lam_R}= \beta(\lam_0) .
\eeq
Denote the fixed point by $\lam_0^*$, and assume\footnote{If $\lam_0^* =0$, we have a so-called Gaussian fixed point
and the formula \rf{s2-2} has to be   modified slightly.} that $\lam_0^* \neq 0$. Since $\lam_0(\xi)$ 
stops changing when $\xi \to \infty$, we have $\beta(\lam_0^*) =0$ and expanding the $\beta$-function 
to first order one finds
\beq\label{s2-2}
\lam_0(\xi) = \lam_0^* + const. \;\xi^{\beta'(\lam_0^*)},~~~
\beta'(\lam_0)= \frac{d \beta}{d \lam_0}.
\eeq
It is seen from \rf{s2-2} that the existence of a UV fixed point implies that $\beta'(\lam_0^*)< 0$.

In a theory like $\phi^4$ in four dimensions it is not clear that there exists a UV fixed point.
The non-existence of such a fixed point will show up in the following way: no matter 
which value of $\lam_R$ we choose, following the curve of constant $\lam_R$ in the 
$(m_0a_0,\lam_0)$-coupling constant plane, the correlation length $\xi$ will never diverge 
along the curve. This implies that there is no continuum limit of the theory with a finite value 
of the renormalized coupling constant.  This seems to be the case for the $\phi^4$-theory in four dimensions
\cite{luscher}. It does not mean that there are no points in the $(m_0a_0,\lam_0)$-coupling 
constant plane with infinite correlation length. In fact, there is an entire curve of such points where 
the lattice model undergoes a second-order phase transition between  the broken and  unbroken 
symmetry\footnote{In the parametrization chosen here, symmetry breaking can occur when we also allow negative 
values of the bare coupling constant $m_0^2$ in \rf{s2-3}. }
$\phi \to -\phi$. However, these points are not related to a UV fixed point, but are related to an infrared
fixed point of the theory. They cannot be reached on a path of constant $\lam_R$ physics and they cannot 
be used to define an interacting quantum $\phi^4$-field theory in the limit where the lattice spacing goes
to zero.

It will be convenient for us to reformulate the above coupling constant flow
in terms of so-called finite-size scaling.  For a regular hypercubic lattice in $d$ dimensions with 
lattice spacing $a$, the physical volume is  $V_d\! =\! N_d a^d$,
where $N_d$ is the total number of hypercubes.
To make sure that $V_d$ can be viewed as constant along  
a trajectory of the kind described above, with $m_R$ and $\lam_R$
kept fixed, we keep the ratio between the linear size  $L\! =\! N_d^{1/d}$ of the lattice 
and the correlation length $\xi$ fixed. In terms of
the renormalized mass $m_R$ and the lattice spacing $a(\xi)$, the ratio 
can also be written as 
\beq\label{s2-6}
\frac{\xi^d}{N_d}= \frac{ 1}{(a(\xi)m_R)^d N_d} = \frac{1}{m_R^d V_d}.
\eeq 
Thus, if we  are moving along a trajectory of constant $m_R$ and  $\lam_R$ in the 
bare $(m_0a_0,\lam_0)$-coupling constant plane and change $N_d$
according to \rf{s2-6}, the finite  continuum volume stays fixed. Assuming that there is a
UV fixed point, such that $a(\xi) \to 0$, we see that $N_d$ goes to infinity even if $V_d$ stays finite, and 
furthermore, again from \rf{s2-6}, that the dependence on the correlation length $\xi$ in \rf{s2-2} can be substituted by a dependence on the linear size $N_d^{1/d}$ in lattice units of the spacetime, leading to
\beq\label{2.2}
\lam_0(N_d) = \lam_0^* + const. \;N_d^{\beta'(\lam_0^*)/d}.  
\eeq
As we saw above, the  absence of a UV fixed point could be deduced by the absence of a divergent 
correlation length along a trajectory of constant physics in the $(m_0a_0,\lam_0)$-plane (i.e.\ a trajectory 
with constant $m_R,\lam_R$). In the finite-size scaling scenario this is restated as 
$N_d$ not going to infinity along any such curve of constant physics.  

We have outlined in this section in some detail how to define and follow lines of
constant physics in the $\phi^4$-lattice scalar field theory, because we want to
apply the same technique to understand the UV behaviour of  lattice theories of quantum gravity.
The most important lesson  is that one is automatically led to UV fixed points (if they exist), if one follows 
trajectories of constant continuum physics. 

\section{CDT}\label{sec3}

\subsection{The lattice gravity program}

Causal Dynamical Triangulations (CDT) represent an attempt to formulate a lattice theory of quantum gravity
(for reviews see \cite{physrep,loll}).
The spirit is precisely that of lattice field theory: one has a continuum field theory with a classical action,
and  defines formally  a quantum theory via the path integral. However, the formal path integral needs to 
be regularized and one way to do this is to use a lattice regularization, where the length of the lattice 
links provides the UV cut-off. The idea is then to search  for a UV fixed point where the lattice spacing $a$ can
be taken to zero while continuum physics is kept fixed, following the same philosophy as outlined above for 
the   $\phi^4$-theory. Immediately a number of issues arise. (1) Given the continuum, classical theory, what 
is a good lattice regularization of this theory? (2) The classical Einstein-Hilbert action is perturbatively 
non-renormalizable. The situation is thus somewhat different from the $\phi^4$-theory in four dimensions. 
The latter exists as a perturbative theory  in $m_R,\lam_R$, the mass and the coupling constant,
and it makes sense to ask whether there exists a  non-perturbatively defined quantum field theory, 
independent of a cut-off for given physical values $m_R, \lam_R$. For a classical action which is non-renormalizable it is 
not clear that the correct way to search for a UV-complete theory is  to keep a lattice version of the 
classical action in the lattice path integral and then search for UV fixed points. (3) What are the physical 
observables in quantum gravity, and how does one stay on a path of constant physics when changing the lattice spacing
in the search for a UV fixed point? Let us discuss these points in turn.

\vspace{12pt}

\noindent
(1) The so-called Regge prescription \cite{regge} provides a way to assign local curvature to piecewise linear geometries 
defined by a ($d$-dimensional) triangulation and the resulting Regge action is a version  
of the Einstein-Hilbert action to be used for piecewise linear geometries. A convenient feature of the Regge formalism is 
its coordinate independence. 
The geometry of the piecewise linear manifold defined by a triangulation is entirely determined
by the lengths of the links and how the $d$-dimensional simplices are glued together. Regge originally 
wanted to use this prescription to approximate a given classical geometry with arbitrary precision
without using coordinates. In the 
path integral we will use it in a different way. We restrict ourselves  to triangulations where all links have the same length
$a$, and then  sum in the path integral over all such triangulations of a given topology, using as our 
lattice action the Regge action for the triangulations. In this way, $a$ becomes a UV cut-off and the hope is that 
this class of piecewise linear geometries can be used to approximate any geometry which would be used 
in the continuum path integral over geometries\footnote{The continuum path integral over four-dimensional 
geometries has not yet been constructed in any mathematically rigorous way, but one expects that the 
geometries will include many ``wild''  geometries which are continuous but nowhere differentiable. In this sense the set of 
piecewise geometries proposed is a set of ``nice'' geometries.}. 

A good analogue is the representation of 
the propagator $G(x,y)$ of a free particle in Euclidean space
 as the path integral over  all paths in $R^d$ from $x$ to $y$, with 
the action being the length of the path. This integral can be approximated by the sum over all paths on a 
hypercubic lattice with lattice links of length $a$. This set of paths is {\it dense} in the set of all continuous 
paths when the distances between paths are measured with the same metric used to define the Wiener measure 
for the set of continuum paths from $x$ to $y$ (see \cite{book} for a detailed discussion with the geometric 
perspective relevant here). We call the way of performing the path integral over geometries\footnote{It should be
emphasized that it is a sum over geometries, not a sum over metrics $g_{ij}$ defining a geometry. In a gauge theory 
this would correspond to a sum over equivalence classes of gauge fields, something one has only been 
able to dream about.}
described above  Dynamical Triangulations (DT) \cite{DT}. The ``proof of principle'' that this method works
is two-dimensional quantum gravity. Seen from a classical gravitational perspective it is a trivial theory since 
the Einstein action in two dimensions is just a topological invariant. 
For a fixed topology the Einstein term does not contribute to 
the path integral, which implies that the action reduces to the cosmological constant times the spacetime volume.
Thus, if we also fix the spacetime volume in the path integral, the action is just a constant and the path integral becomes 
a sum over all geometries of fixed topology and fixed spacetime volume with constant weight. This integral
is still highly non-trivial and  ``maximally quantum'' in the sense that whatever the action is,
in the limit $\hbar \to \infty$ the weight of a configuration in the path integral will be 1. The integral can 
be performed in the continuum, giving rise to  Liouville quantum gravity \cite{liouville,fateev}. At the same time one can also sum 
over the triangulations analytically \cite{2dDT}. One can then verify that in the triangulated case one recovers
the continuum result when the lattice spacing vanishes, $a \to 0$. It is also important to note that the continuum
limit of this  lattice theory is fully diffeomorphism-invariant in the sense that it is identical with 
a diffeomorphism-invariant theory\footnote{No coordinates were introduced at any point in the lattice 
theory, so agreement with a diffeomorphism-invariant theory means that all coordinate-invariant 
quantities which can be calculated agree.}.

While DT works beautifully in two-dimensional spacetime, the generalizations to higher dimensions \cite{higherDT} have 
not been successful yet. The major obstacle has been the nature of the phase diagram of the lattice theory. The goal 
was to find a UV fixed point where one  can define a continuum theory when removing the cut-off.
In our usual understanding this requires a second- or higher-order phase transition. One has found phase transitions
in the bare coupling constants, but so far they have been first-order transitions only \cite{firstorder}  
(see  \cite{firstorder1} for recent attempts to avoid the first-order transitions). This led to the suggestion 
that one should use a somewhat different ensemble of triangulations, 
denoted Causal Dynamical Triangulations (CDT) \cite{originalCDT}.
The difference with the DT ensemble is that one restricts the triangulations to have a global time foliation, which can be viewed as a lattice version of the requirement of global hyperbolicity in classical General Relativity. While 
the DT formalism is inherently Euclidean, one can view the CDT triangulations as originating from triangulations
of geometries with Lorentzian signature. The construction is such that one can   analytically continue  
each individual piecewise linear triangulation to Euclidean signature. In addition,  the associated Regge action also 
transforms as one would na\"{i}vely expect, namely, as $i S [ {\rm LG}]   = -S[{\rm EG}]$, where ``LG'' is the 
Lorentzian geometry and ``EG'' the rotated Euclidean geometry. The path integral is then performed 
over these Euclidean piecewise linear geometries. It turns out that the phase diagram 
of CDT is highly non-trivial and possesses phase transition lines of both first and second order 
\cite{cdtphasediagram1,cdtphasediagram2,nilas}. 
We will provide some details below. It should be emphasized, again with 
the $\phi^4$-example in mind, that the mere existence of a 
second-order line of phase transitions does not ensure that there is a UV fixed point in the theory.

\vspace{12pt}

\noindent
(2) There are at least three ways  to try to 
resolve the problem of the non-renormalizability of  the Einstein-Hilbert action.
One way is to view the theory as an approximation to a larger theory which {\it is}  renormalizable. The Standard Model 
of Particle Physics is the prime example of how this works. Phenomenologically, the weak interactions 
were described by a four-fermion interaction, which is non-renormalizable. However, this is a low-energy 
effective action, which in the Standard Model is resolved into a gauge theory with massive vector particles
(the $W$ and $Z$ particles). Thus, new degrees of freedom were introduced, which made the electroweak theory
renormalizable. Similarly, the effective low-energy theory of strong interactions, involving mesons and hadrons,
was not renormalizable, and again the introduction of new degrees of freedom (the quarks and gluons) made
the theory renormalizable. In the case of gravity, string theory represents such an extension of degrees of 
freedom, but one which is much more drastic than the extensions represented by the Standard Model. And importantly,
while the extension by the Standard Model was dictated by experiments, no string-theoretic extension of gravity has 
yet been forced upon us by experiments. 

Another way to address the non-renormalizability of the Einstein-Hilbert action 
 is to modify the way we view the 
quantum theory in the case of gravity. Loop quantum gravity represents such a route. 
There are still a number of issues that need to be addressed  in this approach,
in particular,  how to obtain ordinary GR in the limit 
where $\hbar \to 0$. We will not discuss this approach any further. 
The lattice regularization of gravity fits naturally into 
the third framework, called {\it asymptotic safety} \cite{weinberg}.  Here one relies on the existence of a non-perturbative 
UV fixed point in some quantum field theory, whose bare Lagrangian can contain many other terms 
in addition to  the Einstein-Hilbert term.  The UV properties of the theory are  defined by this fixed point,
which one should be able to approach  in such a way that the lattice spacing scales to zero,
while keeping a finite number of observables fixed and only adjusting a similar number of bare coupling 
constants.  This is highly non-trivial since  using na\"{i}ve  perturbation theory 
will create an infinite set of new counterterms which cannot be ignored. In the CDT theory
we will look for such UV fixed points by enlarging the Einstein-Hilbert action slightly. It would perhaps be  preferable 
to work with  a more general action, but there are significant numerical limitations which  prevent us from exploring this 
in a systematic way. On the other hand, invoking Occam's razor, CDT quantum gravity in its present 
form is a perfectly  viable candidate theory of quantum gravity, without any compelling reasons to generalize it.
The use of the renormalization group in the continuum provides strong evidence 
for the existence of such a UV fixed point \cite{RG}. However, some truncations  are used to 
obtain these results, whose validity is difficult to assess quantitatively. 
This provides    a strong motivation to search  for such a fixed point 
in lattice quantum gravity, which is an independent way to define quantum gravity non-perturbatively.

\vspace{12pt}

\noindent
(3) One of the steps in the search for a UV fixed point is to choose a suitable set of physical observables
to be kept fixed along the path to the putative UV fixed point. In the case of pure gravity this becomes  non-trivial.
For  the $\phi^4$-theory, one could choose to keep the coupling constants $m_R$ and $\lam_R$ fixed,
because the correlators of the scalar field can be deduced from observations, and the coupling constants can be expressed in terms of these correlators, as mentioned earlier. In a theory of quantum gravity, the concept of a correlator 
as a function of the distance between two spacetime points is problematic, since the distance is itself 
a function of the geometry we are integrating over in the path integral. Thus the concept of a correlation 
length becomes non-trivial, and the whole Wilsonian approach to renormalization -- based 
on having a divergent correlation length on the lattice when one approaches the UV fixed point -- needs 
to be clarified. Even the relation between the UV cut-off (the length $a$ of a lattice link) and any 
actual physically measurable length is not clear a priori. We will return to this in more detail in Sec.\ \ref{sec4}.
 
\subsection{Phase diagram for CDT}    

\begin{figure}[t]
\vspace{-2cm}
\centerline{\scalebox{1.0}{\rotatebox{0}{\includegraphics{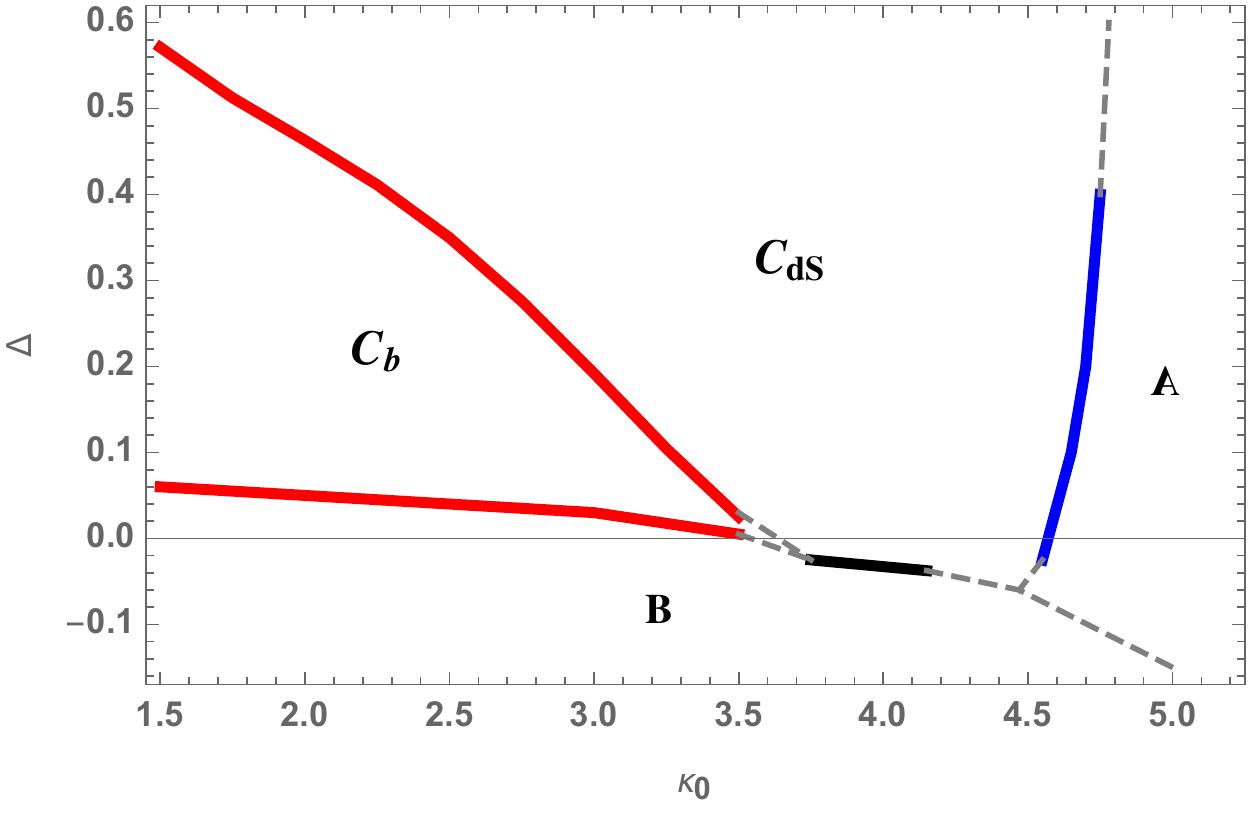}}}}
\caption{The CDT phase diagram. Phase transition between phase $C_{dS}$ and $C_b$ is (most likely) 
second order, as is the transition between $C_b$ and $B$, while the transition between $C_{dS}$ and $A$ 
is first order. The transition between $C_{dS}$ and $B$ is still under investigation, but preliminary results 
suggest a first-order transition.}
\label{fig1}
\end{figure}

In DT and CDT the Regge action for a given piecewise linear geometry 
appearing in the path integral becomes very simple. In dimensionless units, where the lattice 
spacing $a$ is set to 1, the DT Regge action for a four-dimensional triangulation $T$
consisting of $N_4$ four-simplices,  glued together to form a four-dimensional closed manifold in such a way 
that it contains $N_0$ vertices, is given by\footnote{We assume here that $N_0$ and $N_4$ are large, 
since the Euler characteristic of the closed manifold in principle also  appears in \rf{s3-1}, but is ignored.}  
\beq\label{s3-1} 
S[T]=-\kappa_0 N_0(T)+\kappa_4 N_{4}(T).
\end{equation}
In this formula $\kappa_0 \propto a^2/G_0$, where $G_0$ is the bare gravitational 
coupling constant, while $\kappa_4$ is related to the cosmological coupling constant.
Remarkably, no details of the triangulation except for the global quantities $N_4$ and $N_0$ appear in eq.\ \rf{s3-1}.
In the case of CDT we have a foliation in one direction, which we denote the time direction.
The triangulation thus consists of a sequence  of three-dimensional time-slices, where each slice  has the same fixed 
three-dimensional topology (typically that of $S^3$ or $T^3$). 
Each of the time-slices is  triangulated, constructed by gluing 
together equilateral tetrahedra. Neighboring time-slices are joined by four-dimensional simplices,
which come in two types: $(4,1)$-simplices  with four vertices in one time-slice 
and one vertex in one of the neighboring time-slices, and $(3,2)$-simplices, with three vertices in 
one time-slice and two vertices in one of the neighboring time-slices.
The Regge action is slightly more complicated for such a triangulation 
(see \cite{physrep} for a detailed discussion) and has the form
\begin{equation}\label{s3-2}
S[T] =-\left(\kappa_{0}+6\Delta\right)N_{0}(T)+\kappa_{4}\left(N_{4,1}(T)
+N_{3,2}(T)\right)+\Delta\, N_{4,1}(T),
\end{equation}
where $N_{4,1}(T)$ and $N_{3,2}(T)$ denote the number of $(4,1)$- and $(3,2)$-simplices in the triangulation $T$.
For $\Delta = 0$ one recovers the simpler functional form  \rf{s3-1}. Here we will view $\Delta$ as an additional coupling constant\footnote{Originally
in CDT, $\Delta$ was associated with an asymmetry between the lengths of lattice links in the time direction and in 
the other directions.}, with no immediate continuum interpretation.
We thus have the lattice partition function 
\beq\label{s3-3}
Z(\kp_0,\Delta,\kp_4) = \sum_{T} \e^{-S[T]},
\eeq
and the first task is to find the phase diagram in the coupling constant space. We have three 
coupling constants, $\kp_0,\Delta$ and $k_4$. $k_4$ is multiplying the total number of four-simplices and 
acts like a cosmological constant. In the numerical simulations it is convenient to keep the volume $N_4$
of spacetime  fixed.
One can  subsequently perform simulations with different total volumes and study finite-size scaling 
as a function of the total volume, as already mentioned in the discussion of the $\phi^4$-model. Keeping $N_4$ 
fixed implies that we have to fix $k_4$. This leaves 
us with two coupling constants, $\kp_0$ and $\Delta$. In Fig.\ \ref{fig1}
we show the phase diagram of CDT,  determined from Monte Carlo simulations.
The diagram is surprisingly complicated and part of it is still under investigation. 
We refer to the original articles for a careful discussion 
\cite{cdtphasediagram1,cdtphasediagram2,topology1}.
What is important for the present discussion is that in phase $C_{dS}$ in Fig.\ \ref{fig1}, which we denote 
the de Sitter-phase, geometries with continuum-like properties are found. 
Thus, we would like to start with some bare coupling constants
$(\kp_0,\Delta)$ in that phase, calculate the values of some physical observables, 
and then follow the path of constant  physics by changing the bare coupling constants until we 
reach a second-order phase transition point on the phase transition line separating the $C_{dS}$ and $C_b$
phases. If it exists (which is  not at all granted),  this  point will then be a UV fixed point. 

\subsection{Observables and the  UV limit}

What kind of observables can we use in CDT in search of a UV fixed point? We have no fields 
we can associate with lattice vertices or the centers of  (sub)-simplices\footnote{One can in principle associate by hand 
a coordinate system to each simplex, compute transition functions  between the different coordinate systems and
assign metric tensor fields $g_{ij}$ to each simplex, but this  becomes very cumbersome and has  so far not been useful
in a DT or CDT context. It would  also re-introduce a coordinate dependence which is clearly unwanted.}. However, we have geometric 
quantities, like the Regge curvature which is assigned to two-dimensional 
subsimplices in the four-dimensional triangulation,
and we also have the trivial field ``$1(n)$'', which assigns the real number 1 to each four-simplex and 
which turns out to be quite useful\footnote{As observed in \cite{bf},  if one
assumes the existence of a time foliation and  expands the  general continuum effective action for quantum gravity
to quadratic order, one obtains naturally a projection on the constant mode when integrating certain correlators 
over space, as we will do in \rf{s3-5} and as was done in \rf{s2-20} in flat spacetime. In this sense one is naturally led to 
$1(n)$ for such integrated correlators.}. At the same time,  for any given geometry 
we can talk about geodesic  distances between vertices or (sub)-simplices. 
This can be transferred to the quantum gravity theory in the path integral formalism, where one 
 can talk about correlations between some of these quantities when they are separated by a 
certain geodesic distance. The subtlety lies in the fact that this distance has to 
be fixed outside the path integral, since we are integrating over geometries that define 
what we mean by distance. We will return to this point in Sec.\ \ref{sec4}.  Here we will use it in a specific CDT context
where the situation is simpler. 
CDT is special because we have a time foliation, which on the lattice  becomes an explicit time coordinate, namely,
the $n^t$ labelling of the various time-slices. In this sense the set-up in CDT is precisely the lattice set-up 
one would use in proper-time gauge in Ho\v{r}ava-Lifshitz  gravity (HLG) \cite{horava}, although the presence 
of a preferred time in CDT is {\it not} associated with a breaking of four-dimensional diffeomorphism  invariance
(see \cite{loll} for a related discussion). 

Let us introduce the
notation $\la \cO \ra_{N_4}$ for a quantity $\cO$. It signifies  the average of the 
quantity $\cO$, calculated  using the partition function   \rf{s3-3}, but for fixed discrete four-volume $N_4$. 
(In practise the ``calculation'' means that we are performing MC simulations.)
Now we can define the CDT version of \rf{s2-20} for the trivial field $\phi(n) =1$:
\beq\label{s3-4}
N_3(n^t) =\oh  \sum_{\vec{n} (n^t)} (1+1).
\eeq
The notation is as follows: each time-slice is assigned a lattice time $n^t$. 
On this time-slice each three-simplex (tetrahedron) is assigned a label $\vec{n}(n^t)$ by analogy with 
the notation for the hypercubic lattice in eq.\ \rf{s2-20}. This notation is only symbolic, since the three-dimensional 
triangulations are not regular and different time-slices need not contain the same number of 
three-simplices $N_3(n^t)$. 
Each of these three-simplices belongs to precisely two (4,1)-simplices, whose trivial fields ``1'' are
represented in the sum in \rf{s3-4}, and we divide  by 2 to obtain $N_3(n^t)$.
On a regular  lattice, this number is of course trivial and fixed, but here it can vary, as mentioned, and becomes 
a dynamical variable. We now  calculate  averages and correlation functions like in \rf{s2-21}, i.e.\ we calculate
\beq\label{s3-5a}
\la N_3(n^t) \ra_{N_4} 
\eeq
and 
\beq\label{s3-5}
\la N_3 (n^t_1) N_3 (n^t_2) \ra_c = 
\la N_3 (n^t_1) N_3 (n^t_2) \ra_{N_4} -  \la N_3 (n^t_1)\ra_{N_4}  \la N_3 (n^t_2) \ra_{N_4}.
\eeq

\begin{figure}[t]
\vspace{-2cm}
\centerline{\scalebox{1.0}{\rotatebox{0}{\includegraphics{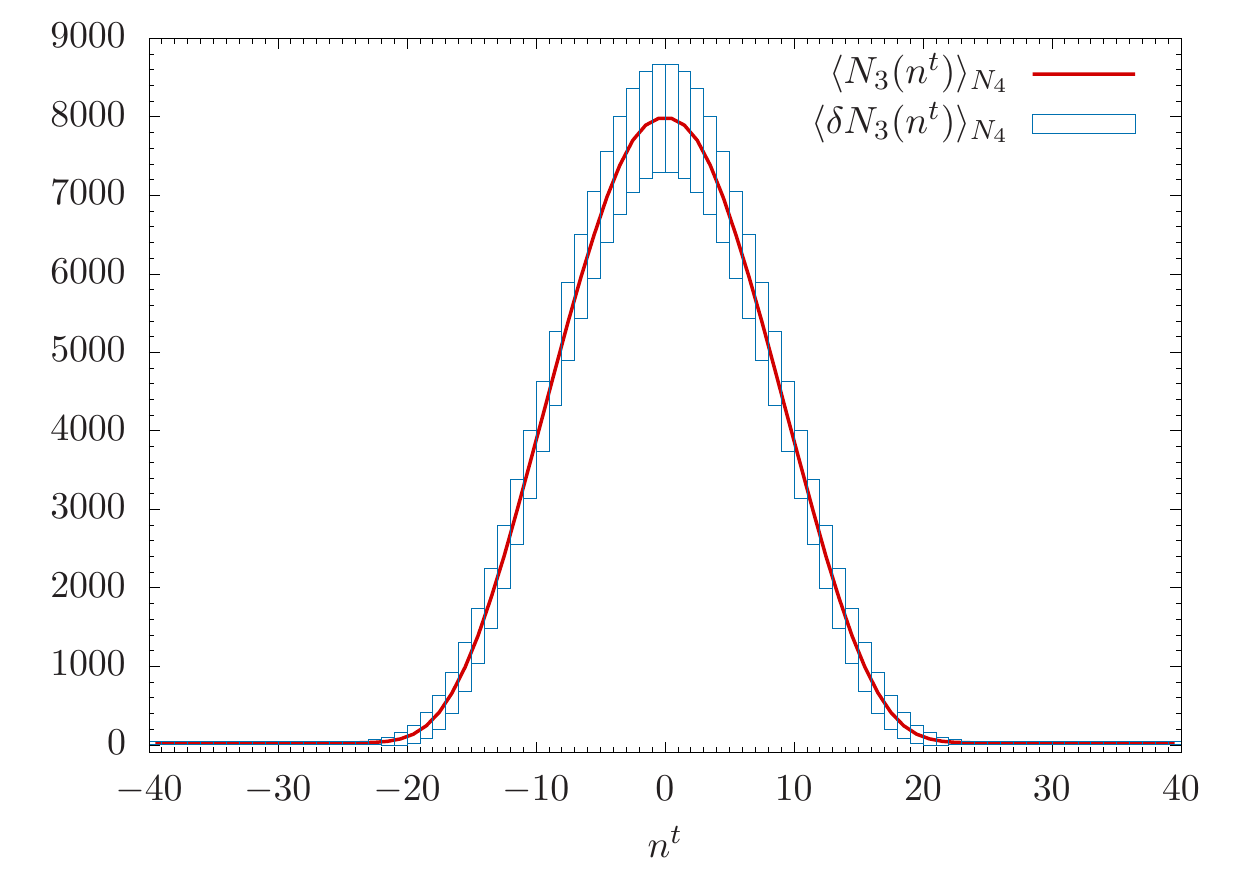}}}}
\caption{The average spatial volume $\langle N_3(n^t)\rangle_{N_4}$ as a result of 
MC measurements for  $N_4=362.000$. 
The  best fit \rf{s3-6} yields indistinguishable curves at given plot resolution.
The bars indicate the average size of quantum fluctuations $\del N_3(n^t)$.}
\label{fig2}
\end{figure}

Fig.\ \ref{fig2} shows the average of $N_3(n^t)$ over many configurations in the case where the topology of the 
spatial slices is that of $S^3$. It also shows the size of the fluctuations, i.e.\ it is a plot of $ \la N_3(n^t) \ra$ 
and $ \del N_3(n^t)= \sqrt{\la N_3^2 (n^t)  \ra_c}$ from \rf{s3-5a} and \rf{s3-5}.
In the region where $ \la N_3(n^t) \ra > 100$, the curve in  Fig.\ \ref{fig2} fits perfectly to the functional form
\beq\label{s3-6}
N_3^{cl}(n^t):= \la N_3(n^t)\ra =  N_4\;\frac {3}{4\om N_4^{1/4}}
\cos^3 \left(\frac{n^t}{\om N_4^{1/4}}\right),
\eeq
where $\om$  depends on $\kp_0$ and $\Delta$.
Despite the fact  that no background geometry enters into the path integral,  a background volume profile 
appears to emerge.
It is identical to  a (Euclidean) de Sitter universe volume profile and
the configurations created by the MC simulations can be viewed as quantum geometries
that fluctuate around this background.
While this is  very interesting\footnote{The dominant ``semiclassical background geometries'' seem to depend on
the topology of space (as do classical solutions of Einstein's equations). If we change the topology 
of space from $S^3$ to $T^3$, the dominant volume profile will  be constant. However, the phase diagram 
is unchanged \cite{topology1,topology}}, our main question here is whether we can use \rf{s3-5a} and 
\rf{s3-5} to follow a path in the bare coupling constant space $(\kp_0,\Delta)$ towards a UV fixed point
in the same way as for the $\phi^4$-theory. More precisely, we want to identify physical observables. Since 
we can perform the MC simulations for various finite volumes $N_4$, we want to use finite-size scaling 
to identify a possible UV fixed point.

A few starting remarks are in order. We have replaced a real field $\phi(n)$ with $1(n)$ in \rf{s3-4} and \rf{s3-5}. Thus we 
cannot {\it necessarily} expect an exponential fall-off and a corresponding correlation length $\xi$ like in  \rf{s2-21}.
However, in the solvable two-dimensional models of both CDT and DT one does find an 
exponential fall-off related to the field $1(n)$ \cite{aw,al}. 
This fall-off is related to the cosmological constants of the models, and the ``mass'' goes
to zero with a vanishing cosmological constant. In four-dimensional gravity we expect massless 
gravitons (and thus maybe no exponential fall-off), but as shown in \cite{bf}, there are  terms 
in an effective continuum action of quantum gravity, which {\it can} lead to such an exponential fall-off, 
e.g.\ the non-local  term
\beq\label{xx1}
 \Gamma^{NL} = - \frac{b^2}{96\pi G}\int d^4x \sqrt{g} \,R \,\frac{1}{\Delta_g^2}\, R,
 \eeq
 where $\Del_g$ is the scalar Laplacian defined in the geometry  related to the metric $g_{ij}(x)$.  
 Expanding the fluctuations to quadratic order  around flat spacetime, $b$ will appear as a mass term.
 We might observe such terms in case of toroidal topology, where the fluctuations we observe seem to 
 be around flat spacetime. If the spatial topology is $S^3$, the contributions from a term like \rf{xx1} would
 mix with contributions from the cosmological term via the curvature of the background geometry used 
 for $S^3$. Thus there might be a number of sources for an exponential fall-off of the (spatial) volume-volume
 correlator.

Eq.\ \rf{s3-6} shows that for fixed $\kp_0$ and $\Delta$ we have a well-defined scaling with $N_4$.
The same is true for the volume-volume correlator, where the MC data
(for  spatial topology  $S^3$) is consistent with the formula
\beq\label{s3-7}
\la N_3 (n^t_1) N_3 (n^t_2) \ra_c =  \gamma^2 N_4 F\Big( \frac{n^t_1}{\om N_4^{1/4}}, \frac{n^t_2}{\om N_4^{1/4}}\Big), 
\eeq
\beq\label{s3-8}
\sqrt{\la N_3^2 (n^t)  \ra_c} = \gamma \sqrt{N_4}\; G\Big( \frac{n^t}{\om N_4^{1/4}}\Big).
\eeq
Here $\gamma$ depends on $\kp_0$ and $\Delta$.  $F$ is some scaling function 
which only depends slightly on $\kp_0$ and $\Delta$, and $G = \sqrt{F}$. 
Unfortunately, we cannot really use eq.\ \rf{s3-7} to extract a correlation length $\xi$ independent 
of $N_4$.  If any $\xi$ could be associated with the correlator, it would already be ``maximal'',
i.e.\ of  order $\om N^{1/4}$, the whole average time-length of the universe, without any fine-tuning 
of the bare coupling constants. 
A condition like \rf{s2-6} then becomes empty\footnote{The situation might be different in the 
case of toroidal spatial topology, where the time extent of the universe is not dynamically adjusted to 
the total four-volume $N_4$. This is presently under investigation.} and we  thus have to 
find other measures to keep continuum physics constant, when  taking the lattice spacing to zero.

Fig.\ \ref{fig2} is for a specific value of $N_4$ and, as remarked above, we already have a scaling 
for fixed values of the bare coupling constants $\kp_0$ and $\Del$. Eqs.\ \rf{s3-6}, \rf{s3-7}
and \rf{s3-8} are these scaling formulas. 
We  see that the height of $ \la N_3(n^t)\ra$ will grow as $N_4^{3/4}$, while the 
fluctuations only grow as $N_4^{1/2}$. For fixed $(\kp_0,\Delta)$ in phase $C_{dS}$, the fluctuations 
will thus decrease relative to  the volume for $N_4 \to \infty$. The interpretation of this is 
that for fixed $\kp_0$ and $\Del$ the limit $N_4 \to \infty$ is one where $V_4 = N_4 a^4$ goes to 
infinity while $a$ stays constant.  

An attempt to replace the $\phi^4$-observables $(m_R,\lam_R)$ with  geometrical  observables is the following.
The {\it physical} volume of spacetime is $V_4 = N_4 a^4$. Similarly, the volume of a time-slice is 
$V_3 (t ) = \la N_3(n^t)\ra a^3$, $t = n^t a$. Let us attempt to take a continuum limit where 
$V_4$ and $V_3(t)$ are {\it finite},  while $N_4 \to \infty$. Such a limit would force $a \to 0$, which is what we want.
How do we ensure that $N_4 \to \infty$ forces $a\to 0$? For the scalar field we had the correlation length 
$\xi$ and $m_R$ which monitored $a(\xi)$. Here we will insist that the relative size of $V_3(t)$ and 
the quantum fluctuations $\del V_3(t)$ stay unchanged if we  scale $N_4 \to \infty$. This would be expected
if $V_3(t)$ can be interpreted as a physical continuum three-volume in the limit $N_4 \to \infty$.
Thus we require that (for sufficiently  large $N_4$)
\beq\label{s3-9}
\ \frac{  \sqrt{(\la N_3^2 (n^t)\ra_c)_{N_4 }}}{\la N_3(n^t)\ra_{N_4}}
= \frac{\del V_3(t) }{V_3(t)}, \quad {\rm independent~of~} N_4 .
\eeq
From \rf{s3-6} and \rf{s3-8} this requirement implies that 
\beq\label{s3-10} 
\om(\kp_0, \Del) \, \gamma(\kp_0,\Del) =  {\it const.}\; N_4^{1/4}.  
\eeq
$\om$ and $\gamma$ are constants independent of $N_4$ for fixed $\kp_0$ and $\Del$. Starting 
out with some $(\kp_0,\Del)$ and a four-volume  $N_4 (0)$ in phase $C_{dS}$, and then increasing $N_4$ will 
force us to change  $(\kp_0,\Del)$ if \rf{s3-10} is to be fulfilled. Continuing to increase $N_4$  will trace out a path in the 
$(\kp_0,\Del)$-coupling constant plane, and the endpoint for $N_4 \to \infty$ will be a candidate for 
a UV fixed point.

The coupling constant flow related to \rf{s3-10} was investigated in \cite{flow} and the conclusion was 
like in the $\phi^4$-case.  There seems  to be no starting point in phase $C_{dS}$ which leads to 
a curve where $N_4 \to \infty$. In fact, while both $\om$ and $\gamma$ change somewhat when 
changing the coupling constants, their product does not change much. We conclude that 
this particular renormalization group 
analysis has not led us to a UV fixed point candidate. But even stronger,  eq.\ \rf{s3-9} 
expresses the simple requirement that if we have a continuum universe of a certain size, it will have quantum
fluctuations of a certain size. However,  our model does not   meet this requirement when 
we relate discretized and continuum variables in  the most natural and simple-minded way .

There are a number of possible interpretations of this result. Firstly, on the technical side, since the 
analysis in \cite{flow} was made, we have obtained a better understanding of the phase diagram. At the 
time of the analysis in \cite{flow}  phase $C_{dS}$ was assumed to extend all the way down to 
phase $B$. Currently the most 
promising phase transition line for a higher-order transition is the $C_{dS}$-$C_b$ transition line,
and   the endpoint of that transition line in particular. We now have a chance to 
approach this fixed point in an easier way  using toroidal spatial topology. This is presently being explored.
Secondly, we may be thinking of 
the quantum universe in the wrong way. In our reasoning we are applying some standard logic related to 
 fluctuations to 
a macroscopic quantity like the three-volume of the universe. Maybe that is wrong. On the other hand, we have 
tried to estimate the size of the quantum universes by constructing the effective action for the three-volume, and comparing with mini-superspace expressions.   The universes are estimated  to have  linear sizes not larger
than 20 Planck lengths \cite{physrep} for the $N_4$-values we are using. Therefore, a picture like that of 
Fig.\ \ref{fig2} should be correct: for a continuum universe of this size we expect significant quantum fluctuations.   
Thirdly, although we tried to emulate  
the flat-space quantum field theoretic way of looking for UV fixed points, 
we have not (yet) been able to identify a divergent correlation length, which is a crucial ingredient of 
the Wilsonian approach to quantum field theory and 
the renormalization group. It is the source of universality and 
dictates the way one moves from the regularized quantum field theory on the lattice to the continuum
quantum field theory. There seems no reason that there should not 
be massless long-range excitations in a theory of gravity related to a universe like ours. 
However, it is much  less clear  what kind of excitations one would observe in a quantum 
universe of the size of 20 Planck lengths, and to what extent one can talk about scaling the lattice 
spacing $a$ to zero compared to the Planck length. The estimates referred to above led to a
lattice spacing of twice  the Planck length. If these estimates can be trusted, our $a$ is far from
sub-Planckian. However, it is possible that the global conformal mode of the metric, whose effective behavior 
we are studying, is not well suited for extracting a correlation length. In other words, it may not be possible to
push the lattice spacing to a sub-Planckian region while maintaining an interpretation that is based on notions which 
are closely related to classical geometry, like ``volume profiles". The question of whether there is a 
correlation length in nonperturbative quantum gravity and whether its divergence relates to a UV phase transition
therefore leads us to an even more basic question:  
what is ``length'' in quantum gravity, when  in the path 
integral one  integrates over the geometries that classically define the length? We turn to 
a discussion of this question in the next Section.

\section{Quantum length}\label{sec4}

 In ordinary quantum field theory,  lengths and distances are defined 
 with respect to a (flat) spacetime metric, which is part of the fixed background structure. One simply has  
 \beq\label{s4-1}
 \la \phi(x) \phi(y)\ra = f( |x-y|),
 \eeq
 where $|x-y|$ is the invariant spacetime distance between the spacetime points $x$ and $y$.
 When trying to define correlation functions like \rf{s4-1} rigorously, e.g.\ on the lattice as in \rf{s2-4}, 
 one may have to rescale fields, coupling constants and the lattice spacing in order to 
 obtain a finite continuum result, but the geodesic distance  $|x-y|$ 
 in (Euclidean) spacetime is  not touched. 
 The situation is similar when we generalize to quantum field theory on a fixed, {\it curved} background. The 
 analogue of the two-point function \rf{s4-1} will still depend on the geodesic distance between $x$ and $y$,
 but also on other coordinate-independent quantities involving the fixed spacetime geometry.
 
 Moving on to  quantum 
 gravity, the path integral will contain an integration over geometries, 
 in addition to the integration over field configurations.
 For these geometries, the geodesic distance between $x$ and $y$ will vary, as will the curvature invariants 
 associated with a given geometry. 
 In the absence of any a priori given background geometry, the only way in which a dependence on a 
 distance (or other geometric invariants) could reappear in some propagator would be with respect to some 
 ``effective" or ``emergent" geometry, generated by the quantum dynamics, and accompanied by 
 quantum fluctuations\footnote{One can of course choose a fictitious ``background''
geometry and expand everything around it. But nothing can depend on this geometry, which 
implies that  distances 
defined with respect to it will be as fictitious as the geometry itself. }. 
The propagator should also reflect this  
to some approximation, depending on the size  the geometric fluctuations. Such an 
``emergence'' of a class of dominant geometries is 
what one observes in the MC simulations of CDT\footnote{To be precise, the emergence of 
classical behavior refers only to those aspects of geometry that are captured by the
behavior in proper time $t$ of the three volume $V(t)$.}  in phase $C_{dS}$. 

For reference, let us examine the situation in 
two-dimensional quantum gravity, which we have argued is 
in some sense maximally ``quantum''. Suppose  we have a universe with the topology of 
a cylinder, where we fix the lengths of the two boundaries to  $L$ and the area (the spacetime volume)
to  $A$. For suitable values of $L$ and $A$ there will be   a ``minimal-area surface'' 
with  constant negative curvature between the two boundaries. Could  this nice, classical geometry
be the one that  dominates the path integral, such that the integration over geometries could  be approximated 
 by considering only small fluctuations around it? It turns out that the answer is no. However, 
{\it if} two-dimensional gravity is coupled to a conformal field theories with a large negative central charge 
the answer is yes. 

Whichever the case may be in four dimensions, some invariant notation of length or distance is clearly needed 
in the quantum theory to construct any meaningful propagators or $n$-point functions.
Again, two-dimensional quantum gravity may provide some guidance for how to proceed. When discussing 
a quantum-gravitational generalization of \rf{s4-1}, we used coordinates $x$ and $y$ to label spacetime 
points, while emphasizing the arbitrariness of this choice. In the context of nonperturbative quantum gravity
it is more convenient to base the construction of invariant correlators on the notion of distance instead.
Thus we integrate only over geometries 
where $x$ and $y$  are separated precisely by a geodesic distance $D$. Equivalently, for a 
given geometry and a given $x$, we average in the matter functional integral over all points $y$ which are separated
a given distance $D$ from $x$, and {\it then}  integrate over all geometries. In this way we obtain a correlation function
$G_\phi (x,D)$, which explicitly depends on what one could call the {\it quantum distance $D$}. Generalizing 
\rf{s2-4}, its definition is 
\beq\label{s4-2}
G_\phi (x,D)  := \frac{\int\hspace{-.5mm} \cD [g_{\mu \nu}] \,\e^{-S[g]} \hspace{-.5mm} 
\int \hspace{-.5mm} \cD_g \phi \,\e^{-S[g,\phi]}
\hspace{-.5mm}  \int \hspace{-.5mm} \d y \sqrt{g(y)} \phi(x) \phi(y) 
\; \del (D_g(x,y) \!-\! D)}{ \int\hspace{-.5mm} \cD [g_{\mu \nu}] \,\e^{-S[g]} \hspace{-.5mm} 
\int \hspace{-.5mm} \cD_g \phi \,\e^{-S[g,\phi]}},
\eeq
where $D_g(x,y)$ denotes the geodesic distance between $x$ and $y$ in the geometry with 
metric $g_{\mu \nu}(x)$.
Even in two-dimensional quantum gravity, the expression \rf{s4-2} is too complicated to compute analytically  
for a scalar field $\phi(x)$. However, for $\phi(x) = 1$ -- the  ``trivial'' field we considered for CDT in Sec.\ \ref{sec3} --
one can in the DT formalism write down a lattice version of \rf{s4-2},  solve 
analytically for the lattice propagator, and take  the continuum limit where the lattice spacing goes to zero
\cite{aw,ajw}. After the continuum limit has been taken  one finds 
\beq\label{s4-3}
G_1(x,D) \propto D^3 F(D/V^{1/4}),\qquad F(x) =1+O(x^4),
\eeq   
if one fixes the spacetime volume to be $V$.
Eq.\ \rf{s4-3} shows that the quantum length $D$ is very ``quantum'', since it has an anomalous dimension,
which moreover  is related to the fractal dimension 4 of the quantum  spacetime. 
If we set $\phi(y) =1$ in \rf{s4-2}, the integral over $y$ is the total volume (in this case the total length) of all points 
at geodesic distance $D$ from $x$, forming a ``spherical  shell''  $S_x (D)$. For a smooth classical 
$d$-dimensional geometry we expect $S_x(D) \propto D^{d-1}$ for $D$ sufficiently small. Here we find instead
\beq\label{s4-4}
G_1(x,D) = \la S_x (D)\ra \propto D^3 \quad {\rm for} \quad D \ll V^{1/4},
\eeq 
which shows that the fractal dimension of two-dimensional Euclidean quantum spacetime is 4.
The important point here is that the distance or length has become a quantum operator, which is natural 
in a theory of quantum geometry. Since  the geodesic distance is a very complicated non-local quantity,  it is 
remarkable that the quantum average of this quantity, defined in eq.\ \rf{s4-2} for $\phi(x) =1$, has 
a non-trivial well-defined scaling dimension.
However, its noncanonical value implies 
a nonstandard scaling behavior of the quantum geodesic distance $D$ in the regularized 
DT-lattice theory  for a  spacetime volume  $V =N_2 a^2$, where  $N_2$ counts the number of 
triangles in the triangulation.  Namely, in a continuum limit
where $V$ stays finite and $N_2 \to \infty$ (and thus $a \to 0$), 
$D$ on average   involves only a number of links $ \propto 1/\sqrt{a}$.
 This is very different from the generic situation in the  $\phi^4$-theory, where
 linear distance in the continuum limit would scale $\propto 1/a$.  In the $\phi^4$-lattice scenario a behavior
 $\propto 1/\sqrt{a}$ would  correspond
 to zero length in the continuum limit. However,  it is  possible and {\it nontrivial} on the DT lattices because of 
 the fractal structure of a generic   triangulation. 

Another related example where distances become quantum comes from bosonic string theory, 
although in a string-theoretical context it is 
usually not presented this way. Bosonic string theory in $d$ dimensions
can be viewed as a theory of two-dimensional quantum gravity with coordinates $(\om_1,\om_2)$ on the world sheet, 
coupled to $d$ scalar fields $X^i(\om_1,\om_2)$,
taking values in the target space $R^d$. Let us consider closed strings, and the so-called tree-amplitude 
for the two-point function. This is calculated by considering two infinitesimal loops 
separated by a distance $D$ in target space, summing in the path integral over all surfaces $X^i(\om_1,\om_2)$
with cylinder topology in target space, with these loops as boundaries, weighted by the string action.
One way to carry out this calculation is to find the classical string solution $X^i_{\rm cl}(\om_1,\om_2)$  
with the given boundaries, perform a split
\beq\label{s4-5}
X^i = X^i_{\rm cl} +X^i_{\rm q}
\eeq
and integrate over the quantum fields $X^i_{\rm q}$. 
Just like in standard quantum field theory, this integration will in general require 
a regularization. In addition, to obtain a finite effective action, $X^i_{\rm cl}$ will need a 
wave function renormalization. However, the distance $D$ 
appears as a parameter in $X^i_{\rm cl}$ and the wave function 
renormalization of $X^i_{\rm cl}$ in reality becomes a renormalization of the distance $D$ in target space,
as shown in detail in \cite{am}.
Like in the case of pure two-dimensional quantum gravity mentioned above, the need for a  renormalization of 
the distance $D$ can be related to a fractal structure, in this case, the fractal structure of the random surfaces 
embedded in $R^d$ \cite{am}. 

The lesson to take away from this discussion is that unless 
some yardstick emerges alongside a ``dominant'' geometry in a non-perturbative path integral over geometries,
or is provided by hand through suitable boundary conditions, a notion of (quantum) distance must be introduced 
in the Planckian regime. As the above examples illustrate, such notions can be found, but will typically behave 
nonclassically or even scale anomalously relative to the volume. Clearly, this needs to be taken into account when 
constructing and interpreting propagators and other geometric observables, for example, in a renormalization 
group analysis near a UV fixed point. Whether such a quantum length possesses a large-scale classical limit
or can be promoted to a ``physical'' length needs to be investigated, and is certainly not a given.

\section{Discussion}\label{sec5}

In the asymptotic safety scenario, quantum gravity is defined as an ordinary quantum field theory  at a
 UV fixed point. We have shown here how one can in principle use computer simulations to 
 search for such a fixed point, in close analogy with the search for a UV  fixed point in a four-dimensional 
 $\phi^4$-theory. The framework of CDT  quantum gravity is  well suited 
 to try and verify the findings of the functional renormalization group analysis in the continuum independently.
 One particular correlation function, that of the spatial volume profile (equivalently, the global conformal mode 
 of the spatial metric), has already been studied extensively, but so far no indication of a UV fixed point has been 
 seen. There could be many reasons for this.
 Despite the compelling evidence  from a body of work in the continuum theory 
 \cite{RG,frank}\footnote{The calculation reported in \cite{frank} seems in particular to be close in spirit
 to the CDT approach.}, such a fixed point
 may not exist, and the asymptotic safety scenario not realized as a way to define quantum gravity beyond 
 perturbation theory. Defining trajectories of constant physics near the Planck scale through an observable that 
 describes the global shape of the universe may be a wrong choice. As emphasized in \cite{flow}, at the very least 
 one would like to repeat the analysis in terms of other, more local observables. A new candidate may be the quantum 
 Ricci curvature \cite{re-ni}, currently under investigation. Our assessment that the lattice version of  $\del V_3(t)$ 
 is too small and does  not increase sufficiently when we move towards the $C_{dS}$-$C_b$ phase transition line
 may be based on our incomplete understanding of how quantum length and volume behave near the Planck scale.
 Another possibility that may be worthwhile exploring is that the quantum-geometric phase transitions in CDT are 
 different from the more conventional Landau-type 
 phase transitions where the Wilsonian renormalization group philosophy works so well.
 In particular, the $C_{dS}$-$C_b$  transition may share traits with the topological phase transitions 
 occurring in condensed matter physics \cite{CM}. The transition is associated with the appearance
 of a localized structure in an otherwise seemingly homogeneous and isotropic universe. It was overlooked 
 for a long time, since the order parameters that exhibit the transition are also of a non-standard nature  with a
 strong topological flavor \cite{nilas}. In addition, one has observed long auto-correlation times in the MC simulations 
 at the  $C_{dS}$-$C_b$ transition, presumably caused by major rearrangements of the internal connectivity of the triangulations in connection with the symmetry breaking. This is again reminiscent of some features seen 
 in topological phase transitions, some of which also have no clear divergent correlation lengths associated with them.
  How to relate such transitions to a UV fixed point in quantum gravity is an interesting challenge.

\section*{Acknowledgement}
JA thanks Frank Saueressig for discussions.
JA  acknowledges support from the Danish Research Council grant {\it Quantum Geometry},
grant 7014-00066B.
JGS acknowledges support from the grant UMO-2016/23/ST2/00289 from the National Science Centre, Poland.
  AG acknowledges 
support by the National Science Centre, Poland, under grant no. 2015/17/D/ST2/03479.
JJ acknowledges support from the National Science Centre, Poland,
grant  2019/33/B/ST2/00589.

\end{document}